# TDOS QUANTUM MECHANICAL VISUAL ANALYSIS FOR SINGLE MOLECULES

P. Contreras*, L. Seijas**, and D. Osorio*
*Physics Department, University of Los Andes, Mérida, 5001, Venezuela.
**Chemistry Department, University of Los Andes, Mérida, 5001, Venezuela.

## ABSTRACT

We have analyzed with pedagogical purposes the relation of the total densities of states (TDOS/PDOS/OPDOS) and molecular orbital diagrams (MOD) for single and isolated molecules of Water and Nitrogen. We use a calculation level HF/6-311G, spreading the energy eigenvalues around the TDOS spikes and using the Mulliken visual analysis in *Multiwfn* to showcase our findings. We discuss the possibility to use these electrons' orbit population diagrams, in conjunction with the more intuitive MOD diagrams to teach the Quantum Mechanics of simple molecular isolated structures.

## INTRODUCTION

The density of states (DOS) is a concept widely used in both the Statistical Physics and the Solid State Physics. The DOS represents the number of energy states available of the system per unit of energy (Reif, 1967). Since in solid-state physics, energy levels in crystals can be represented as a continuum, the DOS can be plotted as a 2D function of the energy. In general, the DOS informs us about the number of levels distributed over an energy interval, it is a measure of how many states are in a small range of energy avoiding the use of the momentum phase space d³$\boldsymbol{p}$.

In solid-state physics, the alternative to use DOS is to express the quantity of interest (electronic specific heat and electronic thermal conductivity among others) in terms of the energy of the system. As some authors point out (Mulhall and Moelter, 2014), each element of volume in the phase space of linear momenta is replaced by a weighting factor in an energy integral, with which it is much easier to work.

The DOS can be written as the sum of infinite number of Dirac delta functions with energies that correspond to the set of eigenvalues $\epsilon_i$ as:

$$D(\epsilon) = \frac{1}{V} \sum_{i}^{\infty} \delta(\epsilon - \epsilon_i)$$

This equation represents essentially "the sum of the total number of different states for all energy levels that electrons are allowed for occupation, that is, the sum of the number of electron states per unit volume, per unit energy".

From a macroscopic point of view, the DOS is explained in courses of Statistical Physics as follows (Reif, 1967): If we define $\Theta(\epsilon)$ as the number of states allowed with energies in the interval between $\epsilon$ and $\epsilon + d\epsilon$ with a macroscopically small $d\epsilon$, we can define the DOS as the proportionality coefficient between the number of macroscopically allowed quantum states and an infinitesimal energy interval such that the derivate rule follows:

$$D(\epsilon) = \frac{d\Theta}{d\epsilon}$$

The previous equation is the most general definition of the DOS and recognizes it as the slope of the number of states $\Theta$ in a $(\epsilon, \Theta)$ plane (Mulhall and Moelter, 2014).

In an isolated molecule in a gas phase without external interactions, however, the energy levels are a few discrete values, so the concept of the DOS as the slope of the number of states becomes questionable. We can argue, for example, that the analysis of the DOS in discrete systems is useless since the energy is not a continuum.

Corresponding author e-mail: pcontreras@ula.ve

However, if we artificially widen the discrete energy levels, it is possible to generate an artificial total density of states (TDOS) even in systems with a few discrete levels. The TDOS formalism exists at the level of research and uses a Mulliken atomic fragments type of analysis (Mulliken, 1955). In this formalism, the TDOS represents a DOS curve for the whole system, PDOS describes a DOS curve for an atom (or fragment), while OPDOS exhibits a DOS curve for a specific space point (i.e. space-resolved among fragments contribution).

The curves obtained this way, can be used to interpret and analyze in a quantum mechanics course, the effects of electronic nature in an isolated molecule such as the relationship bonding/TDOS and the type (character) and amount of atomic orbitals involved in molecular fragments' analysis (Lu and Chen, 2012; Lu, 2017).

On the other hand, bonding is usually represented by molecular orbital diagrams (MOD). The MODs are a qualitative tool to explain chemical bonding in molecules (Tuckerman, 2020. Linear Combination of Atomic Orbitals. New York University, USA. https://chem.libretexts.org) in terms of the linear combination of atomic orbitals (LCAO). Visually, the MODs are diagrams with an increasing function in energy of MO levels, and for different molecular geometries they help to explain the symmetry properties of each molecule (Machado and Faria, 2018), this technique to teach quantum mechanics of simple molecules is still widely used in most textbooks.

Our pedagogical idea is shown in Figure 1. We propose that instead of sketching MOD to students, they can get accustomed immediately with the TDOS diagrams to help describe the nature of bonding in molecular simple systems.

We choose for the TDOS analysis two examples: A water "nonlinear" molecule ($H_2O$) that consists of two atoms of hydrogen linked by covalent bonds to one atom of oxygen. $H_2O$ contains two σ O-H bonds, it belongs to the $C_{2v}$ symmetry point group. Water, despite being a triatomic molecule, with its chemical bonding scheme is difficult to explain many of its bonding properties. For instance, a molecule of water cannot be explained by one unified bonding model.

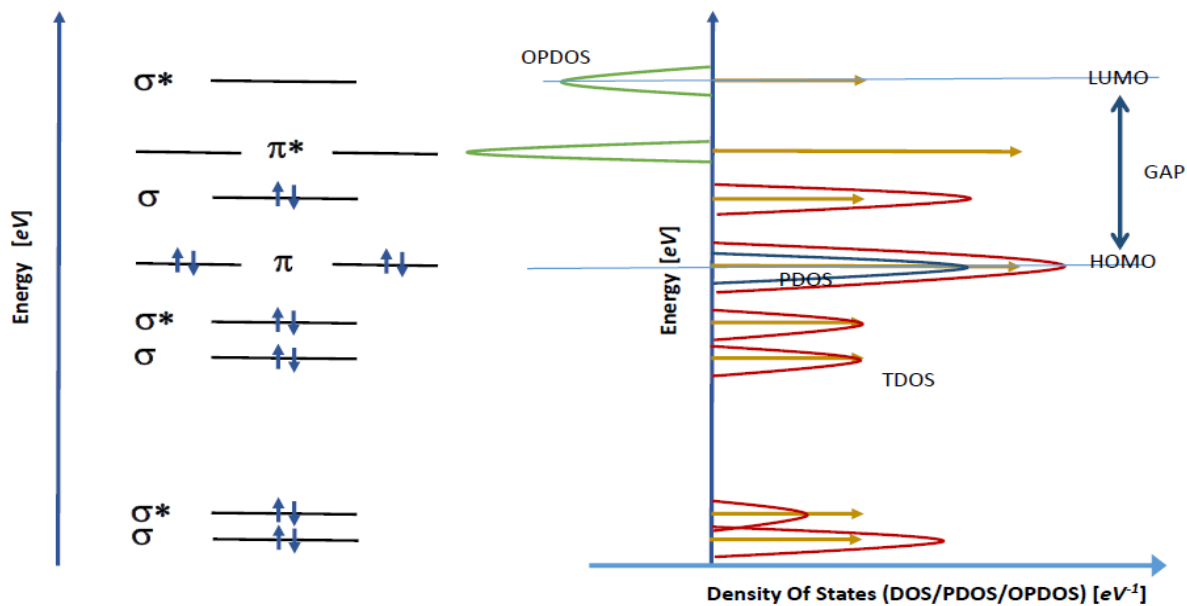

Figure 1. The molecular orbital intuitive diagram (MOD) at the left side and the Eigenenergies' diagram with sketched TDOS/PDOS/OPDOS (right side) and the information attached to this analysis.

$H_2O$ single gas phase molecule has five occupied and three unoccupied molecular orbitals that can be calculated using the Restricted Hartree-Fock wave function HF/6-311G*. Water has a lowest empty molecular orbital (LUMO) formed from the orbital labeled 3a∗, and a highest occupied molecular orbital (HOMO) formed from non-bonding $1b_2$ orbitals (Haas, 2020. Construct SALCs and the molecular orbital diagram for $H_2O$. Saint Mary's College, Notre Dame. https://chem.libretexts.org).

In addition, we add a Nitrogen gas phase molecule (N$_2$) which is a "linear" molecule composed by two atoms with its molecular orbitals labeled by the quantum number $m$. The point group for N$_2$ is D$_{\infty h}$. Nitrogen atoms will form three covalent bonds. In addition, it has a LUMO formed from $\pi_g$* orbitals and an HOMO formed from $\pi_u$ orbitals.

**TDOS theory for a single molecule**

In this section, we define the TDOS (Lu, 2017) as follows:

$$TDOS(\epsilon) = \sum_{i}^{finite\ number} g_{type}(\epsilon - \epsilon_i)$$

If we replace the Dirac function by a broadening function $g_{type}$, we obtain a TDOS spread. The function $g$ can be a Gaussian, Lorenzian, or pseudo-Voight function. In this way, the normalized Gaussian function is defined as follows (Lu, 2017):

$$g_{gauss}(\epsilon - \epsilon_i) = \frac{1}{\sqrt{2\pi}c} e^{\frac{-(\epsilon - \epsilon_i)^2}{2c^2}}$$

where $c$ is proportional to the width at the mean height of the function and corresponds to an adjustable parameter within the definition of the broadened density of states. A larger $c$ implies that the TDOS graph is smoother, which makes the analysis easier to perform. However, this implies that much of the fine structure is masked, within the TDOS.
The Lorenzian function is defined as:

$$g_{lorentz}(\epsilon - \epsilon_i) = \frac{c}{\pi} \frac{1}{(\epsilon - \epsilon_i)^2 + 0.25c^2}$$

with $c$ being the again full width at half maximum, while the pseudo-Voight function $g_{pseudo\ Voight}$ is a linear combination of the Gaussian and Lorenzian functions:

$$g_{pseudo\ Voight}(\epsilon - \epsilon_i) = w_{gauss}\, g_{gauss}(\epsilon - \epsilon_i) + (1 - w_{gauss})\, g_{lorentz}(\epsilon - \epsilon_i)$$

In the latter case, since $g_{gauss}$ and $g_{lorentz}$ are normalized, the normalization condition of $g_{pseudo\ Voight}$ is subject to the value of the weight $w_{gauss}$.

It is convenient to define also the projected partial density of states (PDOS) (Mulliken, 1955; Lu and Chen, 2012; Lu, 2017) as the relative contribution of a particular atom/orbital to the TDOS. The overlap population of the partial density of states (OPDOS) is also a weight function that has large positive values at energies where the interaction between two or more species of atoms is a bonding one in a molecule, and negative values where the interaction is an antibonding one.

These three functions provide valuable information in the analysis of the orbital composition in a MOD, as we will see from the analysis. The PDOS of a molecular fragment A is defined as:

$$PD_A(\epsilon) = \sum_{i} \Xi_A^i\, g_{type}(\epsilon - \epsilon_i)$$

where $\Xi_A^i$ corresponds to the composition of fragment $A$ in the $i$-orbital. The composition of each molecular fragment $\Xi_A^i$ in the $i$-orbital are defined as in Mulliken, that is:

$$\Xi_A^i = (\sum_{a\, \in\, frag\, A} C_{a,i} \sum_{b\, \in\, frag\, A} C_{b,i})\, S_{a,b}$$

While the OPDOS between fragments A and B is defined as:

$$OPD_{A,B}(\epsilon) = \sum_i X^i_{A,B}\, g_{type}(\epsilon - \epsilon_i)$$

where $X^i_{A,B}$ corresponds to the total overlap between fragments $A$ and $B$ in the $i$-orbital and is given by:

$$X^i_{A,B} = 2 \left( \sum_{a\, \in\, frag\, A} C_{a,i} \sum_{b\, \in\, frag\, B} C_{b,i} \right) S_{a,b}$$

where finally, the $S_{a,b}$ term corresponds to the overlap integral among different orbitals. Fragment limits and the total overlap have been defined according to the rules in *Multiwfn manual* section 3.10.3 (Lu, 2017).

**COMPUTATIONAL DETAILS**

For the optimization of the water ($H_2O$) and also nitrogen ($N_2$) molecules, we have used the Gaussian09 program (Frisch *et al*., 2003), with a calculation level HF/6-311G*. The optimization of the structures was carried out using the Berni algorithm. The cutoff values to consider the convergence were set at 0.000450 a.u. and 0.000300 a.u. for the maximum force and the mean square deviation of the force, respectively. To ensure that the structures corresponded to a minimum energy, a calculation of the vibration frequencies was carried out, verifying that there were no imaginary frequencies. The calculation and visualization of TDOS/PDOS/OPDOS as well as the visualizations of the molecular orbitals were carried out using the *Multiwfn program* version 3.3.7 (Lu and Chen, 2012).

**Results for the Water molecule**

To understand what kind of additional information we can get from the artificial TDOS diagram for a single molecule of water (Contreras and Seijas, 2020 https://doi.org/10.13140/RG.2.2.28095.53923), let's consider that for the $H_2O$ calculations at the HF/6-311G* level, the wave function obtained corresponds to the ground state, and the orbitals correspond to the canonical molecular orbitals.

In this case, we have selected fragment 1 (red color) as being the oxygen atom (these correspond to the 2p orbitals), fragment 2 (blue color) has been defined as the hydrogen atoms. The original and expanded densities of states (TDOS, PDOS, and OPDOS) are shown in Figure 2.

In Figure 2, the left and right axes correspond to the TDOS/PDOS and OPDOS curves, respectively. In addition, the vertical dotted line corresponds to the HOMO. The original TDOS is represented by discrete lines, of which the only accessible information corresponds to those of the electronic levels, in which in addition, it is impossible to distinguish the degeneracy of the energy levels. However, if we broaden the energy levels, starting from the height of the TDOS (black line in Figure 1), it is possible to know how dense the energy levels are. For the case of the $H_2O$ molecule, all levels are equally dense as seen in Figure 1.

Furthermore, the broadens that correspond to PDOS (red line for oxygen/fragment 1, blue line for hydrogen/fragment 2) and OPDOS (green line) can identify the characteristics of each bond. For example, the red line approximates the TDOS (black curve) in the region of –0.9 to –0.3 a.u., so we can conclude that the oxygen "2p" orbitals have a significant contribution to the corresponding molecular orbitals of the $H_2O$ molecule.

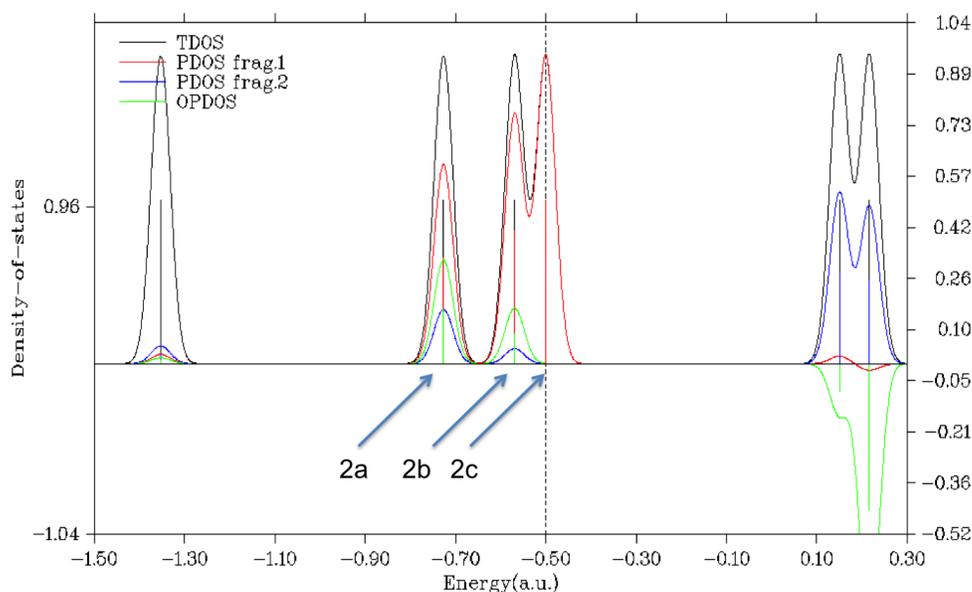

Figure 2. The Eigen-energy levels in the atomic units [a.u.] with the TDOS (black line), PDOS (red line for fragment 1, blue line for fragment 2) and OPDOS (green line). The three blue arrows 2a, 2b, and 2c indicate the energy levels associated with the MO shown in Figure 3 for $H_2O$, right side scale indicates the OPDOS values.

On the other hand, the OPDOS curve (green line) can be greater or less than zero. The height of the OPDOS indicates how favorable or unfavorable the molecular orbital is informing the chemical bond between the "2p" orbitals of oxygen and hydrogen atoms.

Finally, in Figure 2, we can see that all orbitals with energies greater than 0.1 a.u. do not lead to the formation of bonds (green curve negative). Fortunately, these orbitals are unoccupied but if the opposite would have happened, the molecule would break.

Figure 3 shows the MOs associated with the three energy levels indicated in Figure 1 between –0.9 and –0.3 a.u. In Figure 3a, it can be seen that the molecular orbital favors the formation of the chemical bond O-H, since the orbital is located on the oxygen and hydrogen atoms, this orbital corresponds to the highest value of OPDOS (insert 2a in Figure 2).

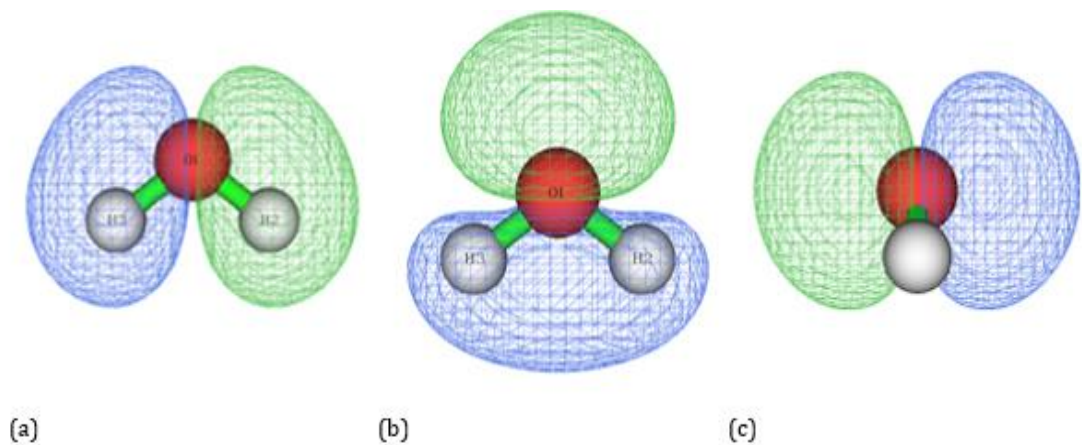

Figure 3. A Multiwfn visualization of the molecular orbitals for a single molecule of water, as are indicated in Figure 2 (plotted from the 2a, 2b, and 2c Mulliken atomic TPDOS fragments' analysis).

In Figure 3b, we can notice that part of the orbital is centered only on the oxygen atom. So, we can conclude that its contribution to the formation of the chemical bond is smaller than in the previous case. This is also observed when examining the height of the OPDOS (the green curve in insert 2b of Figure 2). Finally, the HOMO orbital is centered only on the oxygen atom. So, we can conclude that this orbital practically does not contribute at all to the formation of the chemical bond, at this point the OPDOS value is practically zero. The LUMO for $H_2O$ is not shown in Figure 3.

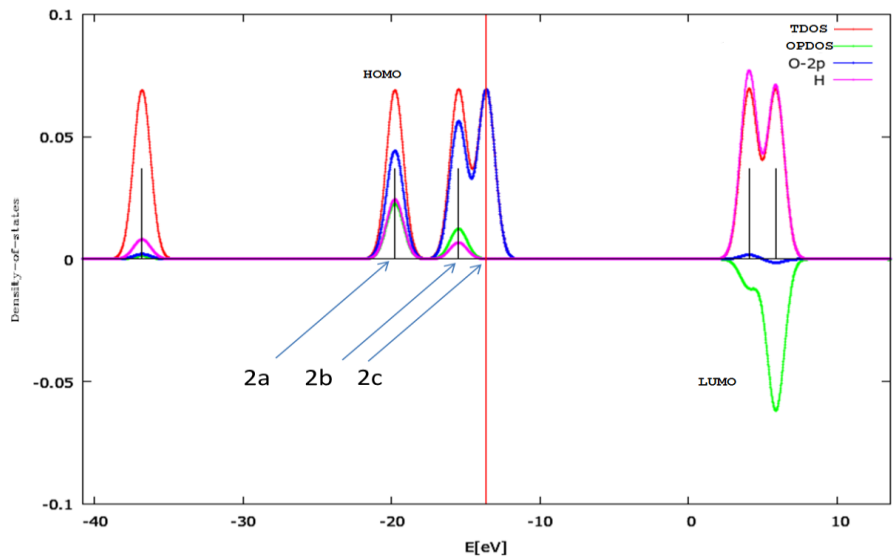

Figure 4. The Eigen-energy levels in electron-volt units for a single H2O nonlinear molecule with the TDOS (black line), PDOS (red line for fragment 1, blue line for fragment 2) and OPDOS (green line) Mulliken analysis. The three blue arrows 2a, 2b, and 2c indicate the energy levels associated with the MO, here we use the same scale for TDOS, PDOS, and OPDOS, therefore, the green OPDOS curve becomes smaller comparing with the one shown in Figure 2.

In Figure 4, we show for $H_2O$ the same results shown in Figure 2 but in eV units, since most times the Eigenvalue energies are expressed in eV and give a better idea of a TDOS analysis.

## Results for the Nitrogen molecule

The second calculation is for a single molecule of $N_2$ in a gas phase. In this case, the wave function obtained corresponds to the ground state of molecular isolated Nitrogen, and the orbitals are the canonical molecular orbitals as well.

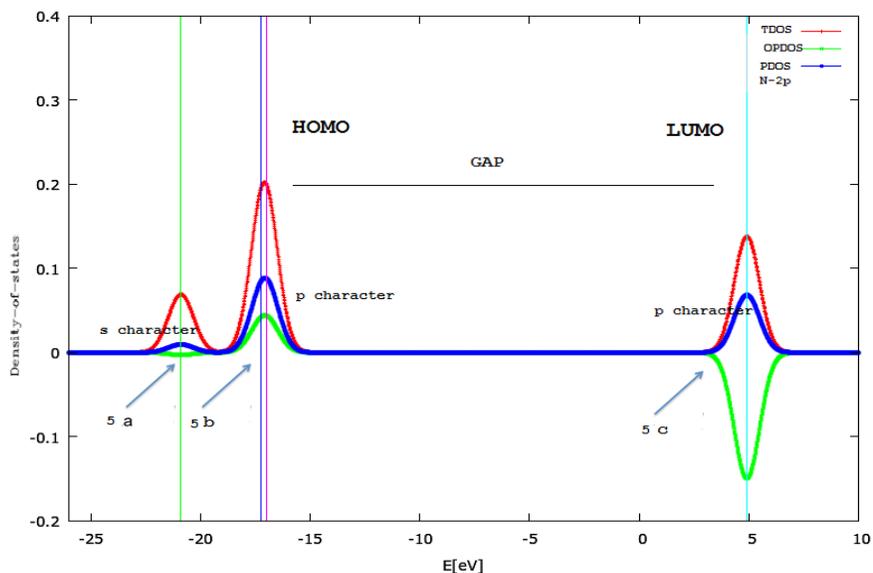

Figure 5. Eigen energy levels in electron volt units for a single gas molecule of $N_2$ with the TDOS (red), PDOS (blue) and OPDOS (green). The arrows 5b and 5c, indicate the energy levels associated with the MO shown in Fig 6 for $N_2$. Arrow 5a points the $\sigma^*(2s)$ orbital.

In Figure 5, the left axes correspond to the TDOS/PDOS and OPDOS curves. In addition, the vertical blue line corresponds to the HOMO $\pi_u(2p_{x/y})$, the vertical cyan line corresponds to the LUMO $\pi^*_g(2p_{x/y})$, and the green vertical line corresponds to the $\sigma^*(2s)$ energy level.

If we broaden as in the previous case the energy levels for $N_2$ starting from the height of the TDOS (the red line in Figure 5), it is possible to know the energy levels' density. The broadens corresponding to the PDOS (blue line for each "2p" orbital) and the OPDOS (green line) identify the characteristics of each bond.

In the case of $N_2$, the levels are not equally dense as in the previous case. The level with a higher density corresponds to an Eigen energy $\epsilon_i$ of –16.19 eV. The lowest Eigenenergy $\epsilon_i$ is –21.43 eV. The maximum level we show is an empty with $\epsilon_i$ 4.99 eV level that corresponds to the LUMO value. The gap value between the HOMO and LUMO is 21.18 eV in a single molecule of $N_2$, according to our calculation.

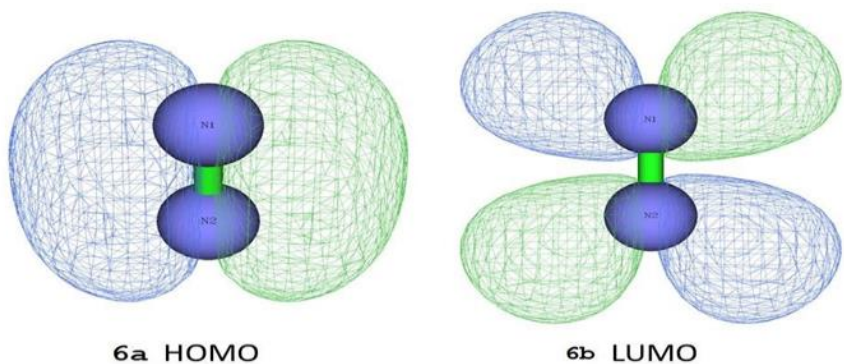

Figure 6. The *Multiwfn* visualization of the molecular orbitals for a single molecule of Nitrogen: (6a) the HOMO and (6b) the LUMO, as are indicated in Figure 5 (plotted from Figures 5b and 5c TDOS fragments' analysis).

.
To complete the analysis, Figure 6 shows the MOs associated with the two p energy levels that define the gap shown in Figure 5 between –17 eV and 5 eV. The $\sigma^*_u$ (2s) MO is not shown in Figure 6. From Figure 6a, it can be seen that the molecular orbital favors the formation of HOMO p N-N bonds. Since the MO are located along both atoms. This MO corresponds to the highest value of the OPDOS. The LUMO antibonding orbital shown in 6b practically does not contribute to the chemical bond, having a negative overlap OPDOS value as seen in the green curve showcased in Figure with the 5c arrow, it is an empty MO.

| Visual Tools | Symmetry picture | Eigenvalues | Atomic fragment analysis | Bonding visualization | Analysis of bond types | Choice of the geometry | Linear - no linear molecule | Shows if levels are occupied? |
|---|---|---|---|---|---|---|---|---|
| TDOS analysis | No | Yes | Yes (PDOS, OPDOS) | Yes | Yes | Yes | Yes | Yes (OPDOS) |
| MOD analysis | Yes (Group Theory) | No | No | No | Yes | Yes | Yes | Yes |

Table 1. The peculiarities of TDOS analysis to understand orbital molecular diagrams respect to the MOD visualization.

In this case, we see that from the TDOS/PDOS/OPDOS information analysis of the energy levels, the Nitrogen "2p" orbitals have a significant contribution to the corresponding molecular orbitals. In addition, the height of the OPDOS here also indicates how favorable or unfavorable the molecular orbital is, informing the chemical bond between the "2p" orbitals in a single molecule of $N_2$.

That is, the overlap population function of the partial density of states (OPDOS) shows that (green curve) is almost zero for the **s-occupied level** shown in Figure 5a, OPDOS is bigger than zero for the most favorable **p-occupied level** shown in Figure 5b, and OPDOS is smaller than zero for the most unfavorable empty **p-high energy level** (Figure 5c).

We finally represent Table 1 that summarizes the TDOS analysis features showcased in this work by comparing with the results of a MOD analysis of several simple single molecular systems (Machado and Faria, 2018).

**CONCLUSION**

From the results previously exposed, we conclude that electronic structure calculations can be suited to perform simple isolated molecules MOD analysis for closed shells. A combined single-molecule MOD, TDOS/PDOS/OPDOS and eigenvalue energy analyses using a Hartree-Fock (HF/6-311G* level) serve a pedagogical purpose in courses of Quantum Mechanics.

In particular, we performed a full fragment DOS analysis for single $H_2O$ and $N_2$ isolated molecules following the information obtained from the HF solutions. The analysis shows illustrative results linking the energy eigenvalues, the localization of the bonds in the MOD, information from the TDOS energy diagrams such as the overlap orbital contribution to each bond, the gap, LUMO and HOMO orbitals with the visualization tool *Multiwfn*, and most important, complement the information given by the MOD intuitive diagrams widely used in the classroom literature.

The same TDOS analysis can be performed for more complicated magnetic molecules at a different *ab initio* level of calculation (UDFT/ B3LYP level) (Burgos *et al*., 2017), it can be done for linear or nonlinear molecules that follow LCAO or symmetry adapted linear combinations (SALCs 2020. Symmetry Adapted Linear Combinations. https://chem.libretexts.org). Table 1 in the previous section summarizes the main findings respect to more popular MOD.

## ACKNOWLEDGEMENTS

P. Contreras acknowledges suggestions by Prof. Belkis Ramirez to improve the manuscript from a previous version. This research did not receive any grant from the University of Los Andes or government agency.

## REFERENCES

Burgos, J., Seijas, L., Contreras, P. and Almeida, R. 2017. On the geometric and magnetic properties of the monomer, dimer and trimer of $NiFe_2O_4$. Journal of Computational Methods in Sciences and Engineering. 17(1):19-28. DOI: https://doi.org/10.3233/JCM-160657

Lu, T. 2017. Multiwfn – A multifunctional wave function analyzer – software manual with abundant tutorials and examples. Version 3.4, sections 3 and 4. Beijing Kein Research Center for Natural Sciences, China. https://web.mit.edu/multiwfn_v3.4/Manual_3.4.pdf

Lu, T. and Chen, F. 2012. Multiwfn: A multifunctional wavefunction analyzer. Journal of Computational Chemistry. 33(5):580-592. DOI: https://doi.org/10.1002/jcc.22885

Frisch, MJ., Trucks, GW., Schlegel, HB., Scuseria, GE., Robb, MA., Cheeseman, JR., Montgomery (Jr.), JA., Vreven, T., Kudin, KN., Burant, JC., Millam, JM., Iyengar, SS. *et al*. 2003. Gaussian 03, Revision B.02. Gaussian, Inc., Pittsburgh PA, USA.

Machado, SP. and Faria, RB. 2018. Explaining the geometry of simple molecules using molecular orbital energy-level diagrams built by using symmetry principles. Quím. Nova. 41(5):587-593. http://dx.doi.org/10.21577/0100-4042.20170198

Mulhall, D. and Moelter, MJ. 2014. Calculating and visualizing the density of states for simple quantum mechanical systems. American Journal of Physics. 82(7):665-673. DOI: https://doi.org/10.1119/1.4867489

Mulliken, RS. 1955. Electronic population analysis on LCAO-MO molecular wave functions. I. The Journal of Chemical Physics. 23(10):1833-1840. DOI: https://doi.org/10.1063/1.1740588

Reif, F. 1966. Statistical Physics. Berkeley Physics Course. McGraw-Hill, New York, USA. Volume 5. pp.398.